\documentclass[conference]{IEEEtran}
\IEEEoverridecommandlockouts
\usepackage{cite}
\usepackage{amsmath,amssymb,amsfonts}
\usepackage{algorithmic}
\usepackage{graphicx}
\usepackage{textcomp}
\usepackage{xcolor}
\usepackage{url}
\usepackage{cleveref}
\def\BibTeX{{\rm B\kern-.05em{\sc i\kern-.025em b}\kern-.08em
    T\kern-.1667em\lower.7ex\hbox{E}\kern-.125emX}}

\begin{document}

\title{A Transfer Learning Based Approach for Classification of COVID-19 and Pneumonia in CT Scan Imaging\\
}

\author{\IEEEauthorblockN{Gargi Desai$^\gamma$, Nelly Elsayed$^\dagger$, Zag Elsayed$^\ddagger$, Murat Ozer$^*$}
	\IEEEauthorblockA{\textit{School of Information Technology} \\
		\textit{University of Cincinnati}\\
		Cincinnati, Ohio, United States \\
		$^\gamma$desaigd@mail.uc.edu, $^\dagger$elsayeny@ucmail.uc.edu, $^\ddagger$elsayezs@ucmail.uc.edu, $^*$ozermm@ucmail.uc.edu}
}

\maketitle

\begin{abstract}
The world is still overwhelmed by the spread of the COVID-19 virus. With over 250 Million infected cases as of November 2021 and affecting 219 countries and territories, the world remains in the pandemic period. Detecting COVID-19 using the deep learning method on CT scan images can play a vital role in assisting medical professionals and decision authorities in controlling the spread of the disease and providing essential support for patients.
The convolution neural network is widely used in the field of large-scale image recognition. The current method of RT-PCR to diagnose COVID-19 is time-consuming and universally limited. This research aims to propose a deep learning-based approach to classify COVID-19 pneumonia patients, bacterial pneumonia, viral pneumonia, and healthy (normal cases). This paper used deep transfer learning to classify the data via Inception-ResNet-V2 neural network architecture. The proposed model has been intentionally simplified to reduce the implementation cost so that it can be easily implemented and used in different geographical areas, especially rural and developing regions.
\end{abstract}

\section{Introduction}
\label{sec:intro}

Emerging and resurfacing bacteria are universal challenges for human health. Coronavirus is a contagious respiratory illness caused by severe acute respiratory syndrome coronavirus 2 (SARS-CoV-2), which is responsible for the COVID-19 pandemic. The first case of novel coronavirus was reported in Wuhan, Hubei province, China, in December 2019~\cite{wu2020new}. The examination of the disease suggests that the outbreak was associated with the seafood market in Wuhan. Coronavirus is a closed Ribonucleic Ac-id (RNA) that is classified among humans, other mammals, and birds. There are six types of coronavirus species that infect humans. Four of these viruses that commonly cause the common cold are 229E, OC43, NL63, and HKU1~\cite{zhu2020novel}. The other two viruses, namely SARS-COV and MERS-COV, are linked to humans' fatal illnesses~\cite{jeffers2006human}. The World Health Organization (WHO) declared the outbreak of COVID-19 pneumonia as a pan-demic in March 2020. Globally, as of 10$^{\mathrm{th}}$ June 2021, there have been 174,061,995 confirmed cases of COVID-19, including 3,758,560 deaths, reported to the World Health Organization (WHO)~\cite{fauci2020covid}. The symptoms of the disease are varied. Common symptoms include respiratory illness, cough, headache, loss of smell and taste, nasal congestion, muscle pain, sore throat, fever, and diarrhea. The presence of the virus in the human body was identified by using sequencing in samples from patients with pneumonia~\cite{zhu2020novel}. The present diagnosis for coronavirus includes reverse-transcription polymerase chain reaction (RT-PCR), real-time RT-PCR (rRT-PCR), and reverse transcription loop-mediated isothermal amplification (RT-LAMP). Other tests also involve nasopharyngeal and oropharyngeal swab tests to detect COVID-19. However, these tests are time-consuming, and the shortage of kits delays the diagnosis~\cite{zhai2020epidemiology}. Other than these tests, computed tomography (CT) scans and chest X-rays (CXR) are used to detect COVID-19 pneumonia. The COVID-19 pandemic is still a primary challenge for the healthcare sector, especially in rural and developing regions with a significant shortage of medical personnel.

Artificial intelligence (AI) attempts to build intelligent models and includes machine learning and deep learning subsets. Deep learning methods can enhance the image features which are not visible in the original image~\cite{raj2020optimal}. Computed tomography of the chest utilizes X-ray apparatus to investigate abnormalities found in tomography tests and to help diagnose the reason for the shortness of breath, fever, chest pain, and other chest symptoms. Computed tomography is accurate, painless, and non-invasive, and it can detect tiny nodules in the lung. Early detection of cases of COVID-19 pneumonia for timely treatment is crucial for avoiding the spread of the epidemic~\cite{ning2020open}. Nonetheless, this remains a difficult task to be completed. A vast number of patients across different locations and the limited medical resources cause a delay in early detection, resulting in delayed decisions on hospitalizations, which increases the chances of cross-infections and poor prognosis. The current RT-PCR test to determine COVID-19 infection has some limitations. The current test (RT-PCR) is not available universally, the processing time can be lengthy, and the sensitivity report varies. As new studies come up, we cannot solely rely on the RT-PCR test for diagnosing COVID-19, especially in patients without symptoms~\cite{harmon2020artificial,xu2020imaging}. It is challenging to distinguish between the CT scans of COVID-19 pneumonia patients and general pneumonia patients because both viruses belong to the same Coronaviridae family. In a cohort of 1014 patients, the ratio of positive chest CT scan results is 88\%, and that of RT-PCR test is 59\%~\cite{zhai2020epidemiology}. Although RT-PCR is most often in agreement, CT can detect COVID-19 in negative RT-PCR tests in people with no symptoms, and it has been an integral part of diagnosis in multiple centers in Wuhan, China, and northern Italy~\cite{valent2021rt}. A recent international expert's report bolsters the use of chest CT for COVID-19 patients with deteriorating respiratory conditions or in a restricted resource environment for medical classification of patients who show average to severe clinical traits~\cite{harmon2020artificial}.

\begin{figure}[t]
	\centering
	\includegraphics[width=0.95\linewidth, height= 3.5cm]{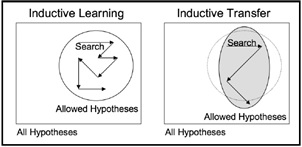}
	\caption{Inductive learning versus inductive transfer methodology.}
	\label{learnings}
\end{figure}

\begin{figure}[t]
	\centering
	\includegraphics[width=0.4\linewidth, height= 6cm]{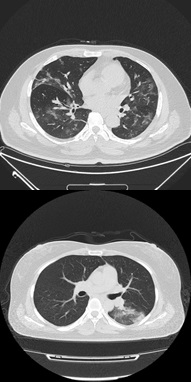}
	\caption{CT scan images showing ground-glass opacities in con-firmed COVID-19 positive patient.}
	\label{ground-glass}
\end{figure}

Given the trend for true artificial general intelligence, transfer learning is something researchers believe can further our progress towards strong AI~\cite{zhu2015towards}. While there might be advanced models, with high precision and knocking all standards, they would be only on very particular datasets and end up with a loss in performance when used in a new task which might still be like the one it was trained on~\cite{zhu2015towards}. This shapes the drive for transfer learning, which goes beyond functions and domains, and tries to leverage knowledge from pre-trained models and use it to solve new problems. In the paper~\cite{pan2009survey}, the authors used domain, task, and marginal probabilities to present a framework for understanding transfer learning~\cite{pan2009survey}. The framework is defined as follows: A domain, D, is defined as a two-element tuple consisting of feature space, $\varphi$, and marginal probability, P($\varphi$), where $\varphi$ is a sample data point~\cite{pan2009survey}. Thus, we can represent the domain mathematically as D = {$\varphi$, P($\varphi$)}~\cite{greenspan2016guest}. Inductive transfer techniques utilize the inductive biases of the initial task to assist the goal task. This can be done in different ways, such as by adjusting the inductive bias of the target task by limiting the model space, narrowing down the hypothesis space, or adjusting the search process itself with the help of knowledge from the original task~\cite{zhu2015towards}. This process is depicted visually in Fig.~\ref{learnings}. 

The main intention of this study is to provide an assisting model for medical personnel to classify CT scans. The proposed research aims to differentiate COVID-19 pneumonia from bacterial pneumonia, viral pneumonia, and normal lungs with no disease from chest CT scans. The model intentionally targets the model design to be simple and accessible to the rural and developing regions where there is an overall shortage of medical personnel, especially lung disease professionals. Moreover, it can be used in any region to support professional medical personnel as a second opinion that can support or discover unseen features in the CT scan so that the medical decision can be revised or request additional tests for the patient to perform the most precise diagnoses.  
In this study, we use the Inception-Res-Net-V2 pre-trained model for transfer learning. In addition, we used publicly available COVID-19 and the pneumonia dataset~\cite{ning2020open} for easy replication of the model. 
\section{Related Work}

Deep learning is a subset of machine learning involved with algorithms stimulated by the shape and role of the brain called artificial neural networks (ANNs). In simple words, deep learning can be thought of as a way to computerize advanced analytics that uses both new and historical data to predict activity. Similar to how humans learn from knowledge and experience gained over time, the deep learning model would perform a task many times, each time adjusting it a little to enhance the output. Deep learning is a significant component of data science, which includes advanced analytics~\cite{tiwari2019evolution,mubarak2021predictive} and predictive modeling~\cite{dias2020deeplms,elsayed2020reduced,mehtab2019robust}. Various techniques are used to create strong deep learning models, including learning rate decay, training from scratch, transfer learning, and dropout. 
To date, deep learning is rising as the foremost machine-learning tool in the general imaging and computer vision domains~\cite{greenspan2016guest}. CNNs have been demonstrated and tested to be powerful tools for a vast extent of computer vision tasks. Studies indicate that medical image analysis groups are quickly adapting to the field of CNNs and other deep learning techniques to a wide variety of applications~\cite{greenspan2016guest}. In medical diagnostics, the assessment of disease relies on both image interpretation and image acquisition. With evolving technology, image attainment has improved substantially over recent years, with improved devices to capture increased-resolution images. However, image interpretation has recently begun to benefit from computer vision technology. Most of the medical image interpretations are performed by physicians. These human-intervened interpretations are limited to subjectivity, with large variations across physicians. Deep learning has proven to be the state-of-the-art basis among many computerized tools, leading to enhanced precision~\cite{van2017fifty}. Deep learning models use networks composed of many convoluted layers that transform the input data (i.e., images/text) to outputs (e.g., lesions present/absent or disease present/absent)~\cite{van2017fifty}.
Deep learning techniques can be used for the early detection of COVID-19 patterns from CT scans of patients. Furthermore, besides the etiological lab confirmation, another primary detection element that assists in identifying COVID-19 pneumonia involves chest tomography imaging and clinical features epidemic~\cite{ning2020open}. As COVID-19 belongs to the Coronaviridae family, it has similar imaging features to SARS-COV and MERS-COV~\cite{zhai2020epidemiology}. As the pandemic progresses and new findings come up, many radiologists report a pattern in the CT scans of patients with COVID-19. The typical features of the CT scan pattern include bilateral, multi-modal, and peripheral ground-glass opacities in the lungs, minor lung lesions, consolidations, ill-defined margins, and light pre-dominance in the light lower lobe of the lungs~\cite{szegedy2017inception,wu2020new}. The CT findings can be seen in asymptomatic patients, and the lesions are evolved into glass opacities and consolidation patterns on the onset of disease~\cite{shi2020radiological}. The ground-glass opacity was seen in 86\% of infected patients, and it involved 76\% bilateral region of the lung in the initial chest CT scans~\cite{zhai2020epidemiology}. Figure~\ref{ground-glass} shows the CT scan images showing ground-glass opacities in confirmed COVID-19 positive patients.
Ning et al.~\cite{ning2020open} proposed a deep learning model using clinical features along with CT scan images. They created open-source clinical data of patients with COVID pneumonia in two cohorts comprising patients with confirmed COVID-19 pneumonia, negative COVID-19 patients, and suspected individuals, and all their CT scan images, clinical features, and lab test results were retrieved. Cohort 1 had data from patients with community-acquired pneumonia, healthy individuals, and negative tested patients for COVID-19. Cohort 2 was used as validation data set with confirmed COVID-19 patients and control cases. The data from cohort 1 and cohort 2 was further divided into type 1 (morbidity outcomes) and type 2 (mortality outcomes) patients. Further, they developed a framework of hybrid learning (HUST-19) using cohort 1 data to diagnose confirmed and suspected cases with mild, regular, severe, and critically ill groups. 
Their hybrid model is comprised of four steps: in step one, they classified individual CT slices into positive CT images, non-informative CT images, and negative CT images using VGG-16 deep learning architecture (six convolutions and two dense layers) with an area under curve (AUC) value of 0.994. In step two, 13 layers of CNN were utilized to change the CT sliced-based prediction to individual patient-based prediction. Using seven deep neural networks, clinical feature-based prediction of patients was implemented, which got an AUC value of 0.978, 0.921, and 0.931 for type 1 and type 2 patients. In the final step, the authors integrated the CT-based and clinical feature-based slices to predict morbidity and mortality results of patients using penalized logistic regression. In addition to the implemented hybrid algorithm by Ning et al.~\cite{ning2020open}, they implemented the Inception-Net-V3 model and Chex Net model using cohort 1 data. The outcomes show that all the models used have similar accuracy in predicting the morbidity and mortality results. 
Another work by Sharma et al.~\cite{sharma2020drawing} classifies CT scans between COVID-19 pneumonia and other viral pneumonia using machine learning~\cite{sharma2020drawing}. The system is built on the residual neural network architecture of Microsoft Azure. The model contained 800 CT images of COVID-19 infected patients, 800 CT images of healthy people, and 600 CT images of other viral pneumonia, which gave overall accuracy of almost 91\% with some false indications. This article also discusses whether the Chest CT scan can be the alternative test for the current RT-PCR screening test. Liu et al.~\cite{liu2020differentiating} proposed a machine learning framework to differentiate COVID-19 pneumonia from general pneumonia. Prior to the feature extraction being performed, their framework implemented a region of interest (ROI) delineation founded on the glass opacities on 73 confirmed COVID patients and 27 general cases of pneumonia. The relief algorithm was used to select features from the 34 extracted features. The authors classified the features using the ensemble of the bagged tree (EBT) algorithm along with logistic regression, SVM, decision tree, and KNN with Minkowski distance~\cite{liu2020differentiating}. Their method yielded 88.6\% sensitivity, with EBT giving all the classifiers consistent performance. They stated that the study could be improvised with more amount of data. Another study by Lin Li et al.~\cite{li2020using} developed a 3D deep learning model referred to as COV-NET for diagnosing COVID-19 using volumetric chest CT. The study in~\cite{li2020using} included CT scans of community-acquired pneumonia and other non-pneumonia abnormalities. The deep learning model was developed on ResNet architecture and gained a sensitivity of 90\%. Initially, the region of interest is obtained by using U-net~\cite{ronneberger2015u}. The model first extracts features combined through max-pooling operations before feeding into the SoftMax activation function~\cite{goodfellow2016deep} to generate a probability score for classifying each type (COVID-19, Community-acquired pneumonia, and non-pneumonia). Multinational data was collected, and multiple classifications were performed to increase the generalizability of detecting COVID-19 using CT scan images~\cite{harmon2020artificial}. The authors developed hybrid 3D and full 3D models based on Densenet-121 architecture, giving them up to 90.8\% accuracy. Initially, the lung segmentation model was developed using AH-Net architecture, followed by image classification models. Only a 10\% false-positive rate was noticed in 140 patients with confirmed COVID pneumonia and other pneumonia.

There are several other studies that show the capabilities of deep learning on chest radiographs and image classification. For example, in the paper by Ningwei Wang, transfer learning models ResNet-101 and ResNet-152 are used on chest X-Ray images to classify COVID-19, normal and viral pneumonia. The study in~\cite{wang2020deep} states that the model can achieve 96.1\% accuracy on the test set. Encouraged by these research papers, we aim to further examine the deep learning model Inception-ResNet-V2 and evaluate its feasibility for COVID-19 diagnosis. The motivation behind this study is to assess the performance of the available pre-trained model, to classify COVID-19 pneumonia from CT scan images. In addition, we intentionally tried to find an economic model that can be used to classify COVID-19 pneumonia in different areas and regions, especially the developing countries and rural areas that can not afford cost expensive labs. Additionally, this paper investigates the plausibility of top-tier pre-trained neural networks for computer vision. 

\begin{figure}[t]
	\centering
	\includegraphics[width=0.95\linewidth, height= 3.5cm]{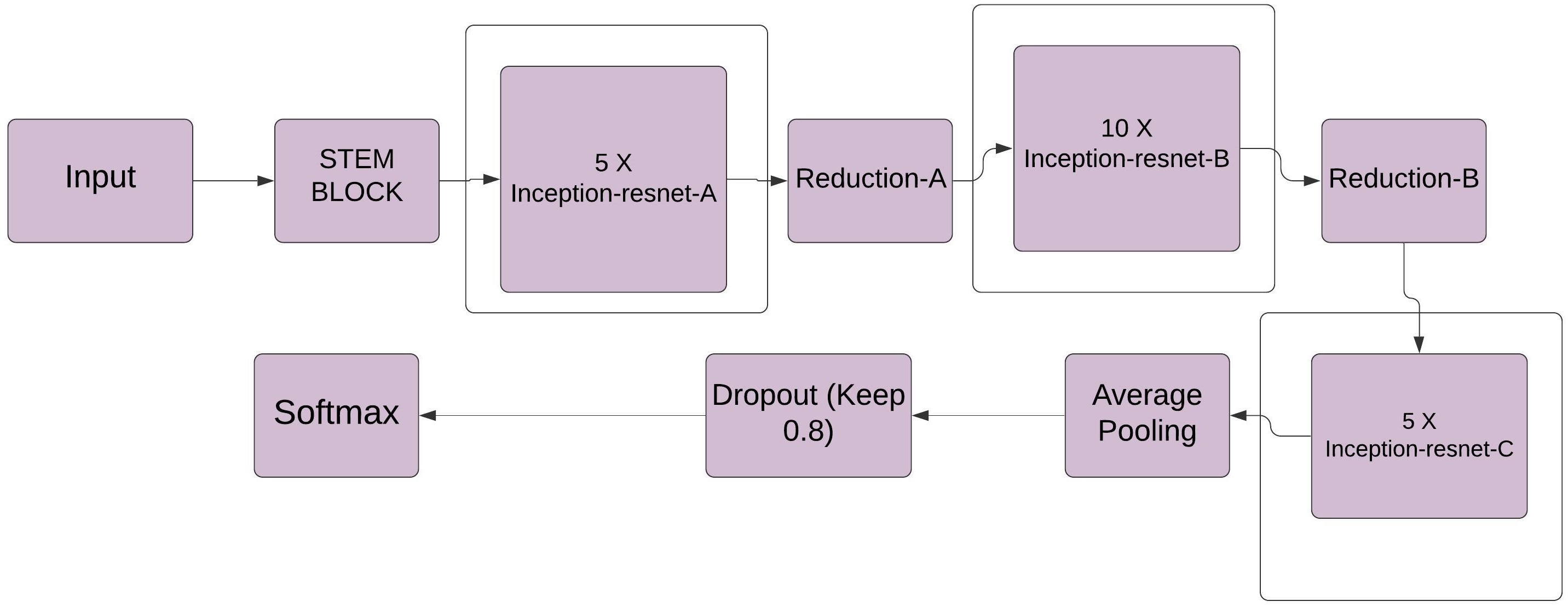}
	\caption{Inception-ResNet-V2 architecture.}
	\label{inseptionV2}
\end{figure}

\section{Model Architecture}

\begin{figure}[t]
	\centering
	\includegraphics[width=0.95\linewidth, height= 3.5cm]{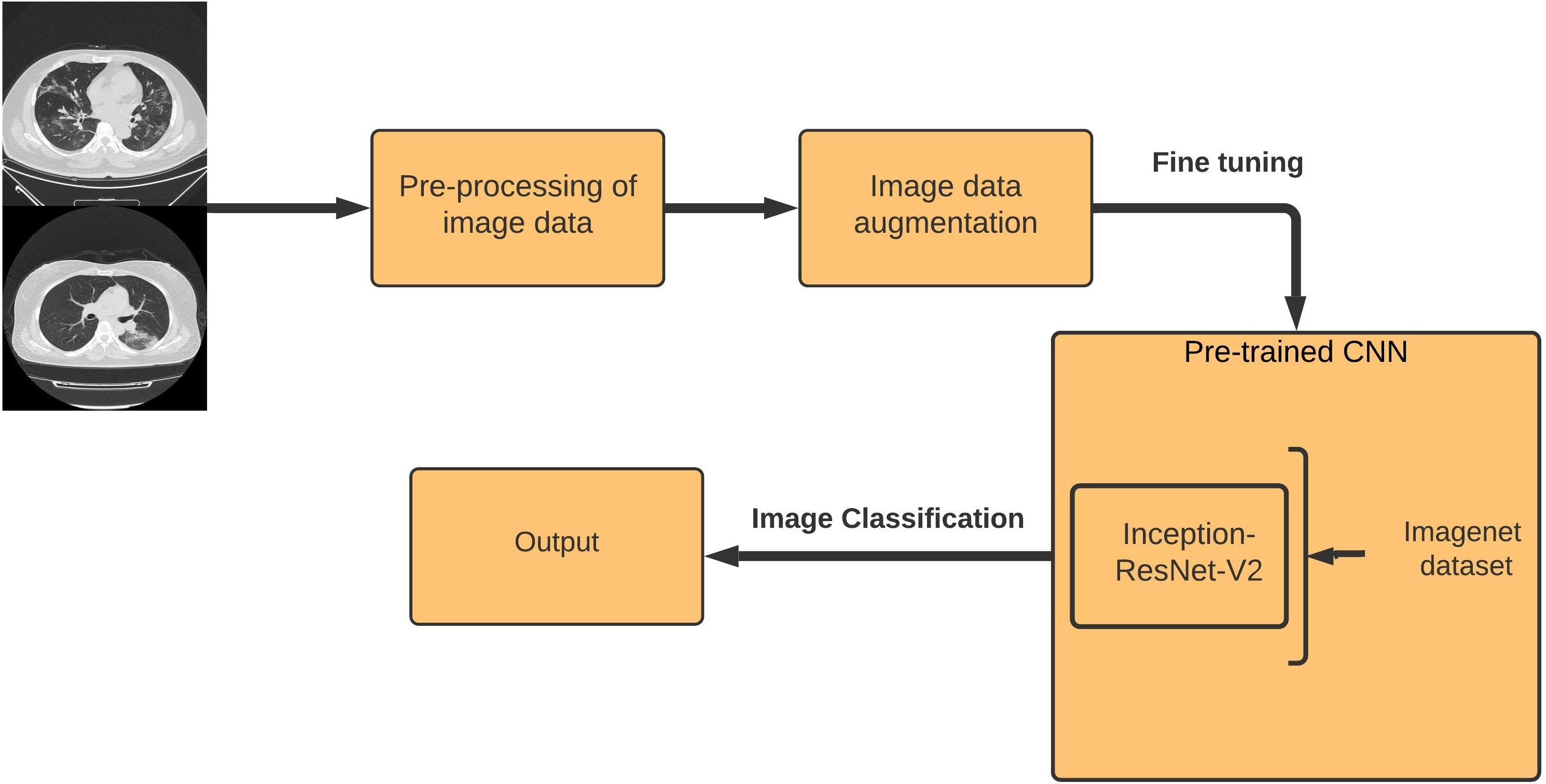}
	\caption{Framework of the proposed system approach.}
	\label{framework}
\end{figure}

The proposed model includes three main steps: pre-processing, fine-tuning Inception-ResNet-V2 CNN, and classification stages. Figure~\ref{framework} shows the proposed model approach.

\subsection{Data pre-processing}

One of the curial challenges researchers face in the medical field is the limitation of datasets availability. In this research, we have used integrative CT images for the COVID-19 dataset, including 1521 patients with or without COVID-19 pneumonia and their recorded clinical features epidemic~\cite{ning2020open}. The dataset is an open-source dataset and is available for academic research. Another open-source dataset containing CT scan images of different types of pneumonia, including COVID-19, non-COVID-19, viral pneumonia, bacterial pneumonia, and healthy lungs, is utilized in this study~\cite{yan_dataset}. The database consists of 11931 CT scan images comprising 4698 CT images of normal lungs with no disease, 4001 CT images of confirmed COVID-19 Pneumonia, 1255 CT images of Viral Pneumonia, and 1977 images of Bacterial Pneumonia. 

\subsubsection{Data preprocessing stage}
To obtain a proper input image to the model, all the images are pre-processed before feeding to the network. All the CT scan images are in JPEG format. Further, all the JPEG images are decoded and resized to 224$\times$224$\times$3 pixels to fit the model. Next, all the JPEG converted images are arranged based on image classes. For the experiments, the data is divided into 60\% training and 20\% validation, and 20\% test sets with an equal number of image ratios of all classes. 
The test set is divided into three subsets with an equal number of image ratios in all classes.

\subsubsection{Data augmentation}
Image augmentation is usually required to improve the performance of deep networks. Image augmentation artificially creates training images through alternative ways of processing or a combination of multiple processing, like random rotation, shifts, shear, flips, etc. An image data generator is defined to train the models at modified versions of the images, such as at different angles, flips, rotations, or shifts. As the data is limited, image augmentation is used to expand the data for the experiments. For the binary classification experiment (COVID-19 versus Normal Lungs), rotation range, width shift range, height shift range, and horizontal flip augmentation techniques are used. For multiclass classification experiments (all four pneumonia classes), width shift range, height shift range, shear range, zoom range, rescale, and horizontal flip augmentation methods are used.

\subsection{Transfer learning and Inception-ResNet-V2}

Transfer learning is a time-saving method in computer vision to build an accurate model. Instead of starting to develop a model from an ideal model, one can use the pattern learned when resolving distinct problems~\cite{zhu2015towards}. In addition, it provides the transfer of the prior learning experience of the model into the current model architecture. Moreover, transfer learning can help to design models using a lower computing processing budget compared to similar non-transfer learning-based models.

Inception-ResNet-V2 is a CNN trained on a vast set of images from the ImageNet database~\cite{szegedy2017inception}. Inception-ResNet-V2 combines the Inception architecture with residual connections. The network is 164 layers deep and can classify images into different class categories. Therefore, CNN has learned full feature depiction for a wide range of images. The network has an image input size of 299$\times$299, and the output is a list of estimated class probabilities. This architecture is based on a combination of the previous Inception model structure and residual connection. In the Inception-ResNet-V2 block, multiple-sized convolutional filters are combined with residual connec-tions~\cite{szegedy2017inception}. The usage of residual connections not only avoids the degradation problem caused by deep structures but also reduces the training time~\cite{nguyen2018deep}. Figure~\ref{inseptionV2} shows the basic network architecture of Inception-ResNet-V2 where:
\begin{enumerate}
	\item \textbf{Inception blocks:} The pre-trained model has a base convolution base layer and classifier. A few essential, enhanced convolution blocks in the Inception-ResNet-V2 model are mentioned below. 
\item \textbf{Residual Inception blocks:} Each Inception block is followed by a filter expansion layer (1$\times$1 convolution without activation) which is used for scaling up the dimensionality of the filter (kernel) bank before the addition to match the depth of the input. The difference seen shows that in Inception-ResNet, batch-normalization is employed only on top of the normal layers but not on top of the summations. 
\item \textbf{Scaling of Residuals:} As stated in the paper, if the number of filters exceeded 1000, the residual variants started to exhibit instabilities, and the network has just "died" early in the training stage, meaning that the last layer before the average pooling began to produce only zeros after a couple of tens of thousands of iterations~\cite{szegedy2017inception}. This could not be prevented by lowering the learning rate or adding an additional batch normalization to the presentation layer. According to them, cutting down the residuals before adding them to the previous layer activation appeared to stabilize the training~\cite{szegedy2017inception}. To scale the residuals, scaling factors between 0.1 and 0.3 were selected. 
\end{enumerate}
This paper uses transfer learning with Inception-ResNet-V2 model architecture as the backbone.

\section{Experiment and Result Analyses}

In this paper, we designed two architectures: one to classify COVID-19 pneumonia from normal lungs, and the second is to classify COVID-19 pneumonia, bacterial pneumonia, viral pneumonia, and normal lungs. We named the first model binary classification and the second multiclass classification task to simplify tracking these two experiments.

For the binary classification experiment, the dataset is balanced with a total of CT images to be 8699. For this model, the dataset is divided into 60\% of images for training and 20\% images for validation, and 20\% of images for testing the accuracy of the model. To evaluate the results precisely, the test set is further divided into three subsets. All the CT images are in JPEG format and are resized to a fixed size before they can be fed to the deep learning models for training. The CT images are resized to 224$\times$224 pixels, which is ideal for the Inception-ResNet-V2 model. The images are augmented. Four custom layers are added in the beginning to the pre-trained models so that they can be trained on the dataset.
Further, a Flatten layer to flatten the features, a batch normalization layer with kernel-regularization, two fully connected convolution layers with ReLU activation, and a dropout layer to overcome overfitting is added. Finally, a dense output layer using the SoftMax function as the activation function is added to the model. Since the initial half of the model is already pre-trained, these layers have been set to freeze (i.e., the trainable attribute of the initial half of the model is set to "False"). This helps to accelerate the training process. The proposed model is compiled with the Adam optimizer~\cite{kingma2014adam} with a learning rate 0.001, $\beta_{1}$ = 0.9, $\beta_{2}$=0.999, and $\varepsilon$ = $1e^{-07}$. The categorical crossentropy is used as the loss function.

For multiclass classification, the data is split into a 60\% training set, 20\% validation set, and 20\% testing set with an equal ratio of all classes. The dataset has a total of 11931 CT scan images. The input layer has 224$\times$224 image size. Image augmentation is applied as discussed. Furthermore, to overcome the imbalance of data in the experiment, class weight is determined.
The batch size is set to 32. To improve the performance, custom layers like flatten and batch normalization layers with kernel regularization, two fully convoluted dense layers with the rectified linear unit (ReLU) activation~\cite{teh2001rate,elsayed2019effects}, and a dropout layer is added to fine-tune the model according to the dataset. Finally, a dense output layer using the SoftMax function as the activation function is added to the model. The architecture is trained with Adam optimizer and uses categorical crossentropy as a loss function~\cite{kingma2014adam}. We have used the early stopping callback function to stop training when a monitored quantity has stopped improving.
Further, we also apply the Reduce-LR-On-Plateau callback function to reduce the learning rate when a metric has stopped improving~\cite{krzyston2022neural}. Models often benefit from reducing the learning rate by a factor of 2:10 once learning stagnates. Reduce-LR-On-Plateau callback monitors a quantity, and if no improvement is seen for a 'patience' number of epochs, the learning rate is reduced.

\subsection{Environment Setup}
The fine-tuned Inception-ResNet-V2 pre-trained CNN architecture is applied for the classification and detection of COVID-19 pneumonia. The final fully convoluted (FC) layer is used for the classification in both experiments. Python 3.7.10 is used as the programming language to train the CNN model with TensorFlow library version 2.0. The whole experimentation is performed on a Google Collaboratory version: 10.1 with Tesla P100.
To evaluate the proposed work, the model's effectiveness is calculated based on metrics results conducting the experiments and explores the fine-tuning technique of transfer learning by extracting the features of pre-trained CNN networks. The experimental study is performed on publicly available datasets. To estimate the accuracy of the models and get insights into their performance, we calculate the precision, recall, F1-score, and confusion matrix metrics. 




\subsection{Result Analysis}


\subsubsection{Binary classification}
In this classification, COVID-19, from normal lungs, has been detected with an average accuracy of 90.85\%, a sensitivity of 0.97, a specificity of 0.85, and an F1-score of 0.91. The sensitivity rate of the experiments shows us the CT scan detected as positive for COVID-19. The model outperforms other study models in the binary classification of COVID-19 CT against the CT of normal lungs with no disease. We have compared the models later in this section. The confusion matrix gives detailed insights into the test results of the model. Compared with lab tests which can take up to two days, and there is a high possibility of false negatives, this research can help give a rapid and accurate diagnosis. The proposed model's high sensitivity can provide physicians with a 'second opinion' in cases where the RT-PCR is negative, which can be common. It is something that should be considered on a broader scale. Detailed test results for each of the three test sets have been shown in Table~\ref{binaryClassificationTable}.


\begin{table}
	\centering
		\caption{Evaluation results of binary classification using our proposed model.}
	\footnotesize
	\begin{tabular}{@{}l|lllll@{}}
		\hline 
		
		\multicolumn{1}{l|}{}&\multicolumn{1}{l}{\textbf{Metrics}}&\multicolumn{1}{l}{\textbf{Test1}}&\multicolumn{1}{l}{\textbf{Test2}}&\multicolumn{1}{l}{\textbf{Test3}}&\multicolumn{1}{l}{\textbf{Avg.}}\\
		\hline 
		\textbf{COVID-19} & Precision& 0.87&0.82&0.85&0.85 \\
		\textbf{Pneumonia }& Recall& 0.92&1.00&1.00&0.97\\
		& F1-Score& 0.90&0.90&0.92&0.91\\
		& Specifity& 0.88&0.82&0.85&0.85\\
		\hline
		\textbf{Normal}& Precision& 0.93&1.00&1.00&0.98\\
		\textbf{Lungs}	& Recall& 0.88&0.82&0.85&0.85\\
		& F1-Score& 0.91&0.90&0.92&0.91\\
		& Specifity& 0.92&1.00&1.00&0.97\\
		\hline
		\textbf{Overall} & Accuracy&90.33\%&	90.16\%	&92.06\%&	90.85\%\\
		\hline 
	\end{tabular}

	\label{binaryClassificationTable}
\end{table}

\begin{figure}[t]
	\centering
	\includegraphics[width=0.95\linewidth, height=3cm]{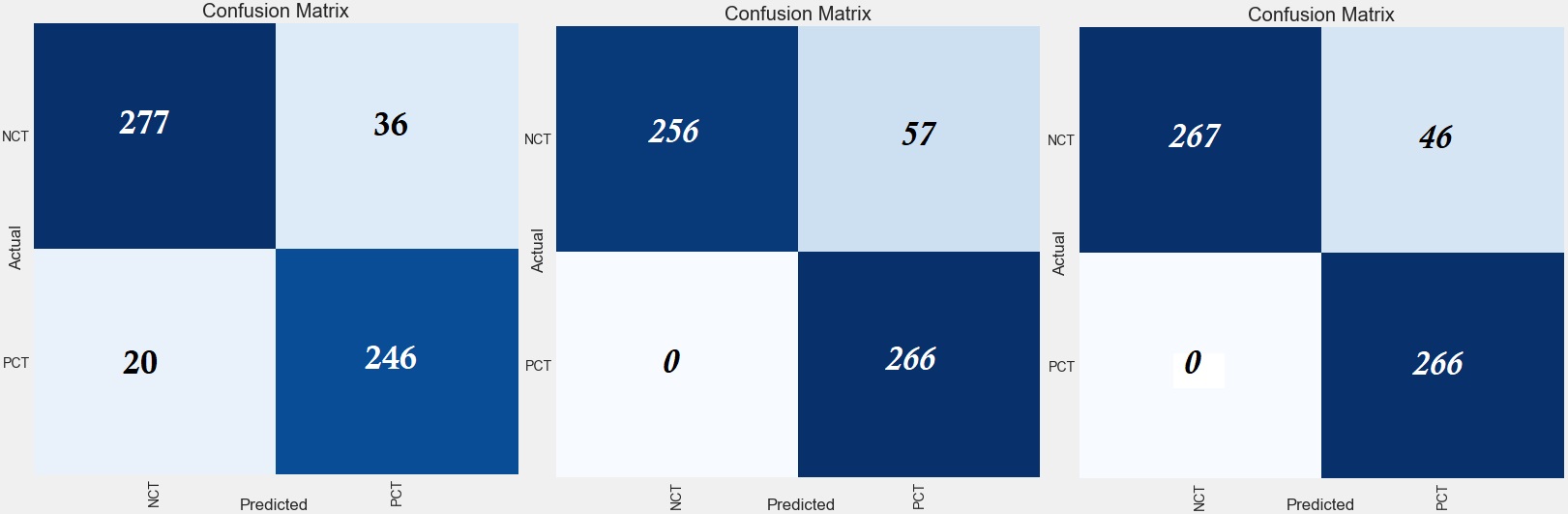}
	\caption{Confusion Matrix of Binary Classification on three test sets [PCT: Positive COVID-19 CT, NCT: Normal Lungs CT].}
	\label{confusionMatrixBinary}
\end{figure}

\begin{figure}[t]
	\centering
	\includegraphics[width=0.96\linewidth, height= 3.5cm]{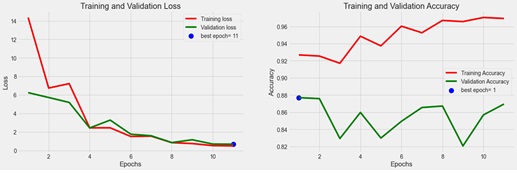}
	\caption{Training and validation loss and accuracy in the COVID-19 binary classification task.}
	\label{trainvsvalidationBinary}
\end{figure}

The efficiency of the implementation is seen in the confusion matrix of three different test sets, as shown in Figure~\ref{confusionMatrixBinary}. In the first test set, the confusion matrix shows that the PCT (Positive COVID-19 CT) is misclassified 20 times and correctly classifies COVID-19 246 times. In the second test set, the confusion matrix shows that the PCT is never misclassified, and the same is seen in the third test set. For binary classification, the dataset achieves an average accuracy of 90.85\%. The classification report of the first test set gives a sensitivity rate of 0.92 for PCT classification and 0.88 for normal lung CT scans. The sensitivity rate of the second test set is 1.00 for PCT classification and 0.82 for normal lung CT scans. Finally, the third test set shows the sensitivity rate is 1.00 for PCT classification and 0.85 for normal lung CT scans. The F1-score for binary classification was 0.91. 
The training and validation loss and accuracy graphs are shown in Fig.~\ref{trainvsvalidationBinary}. The early callback function stops the training on epoch 11. It shows that validation loss is low and slightly higher than the training error, indicating that the model is a good fit for the data.

\begin{table}
		\caption{Evaluation results of multiclass classification using our proposed model.}
	\centering
	\footnotesize
	\begin{tabular}{@{}l|lllll@{}}
		\hline 
		
		&\multicolumn{1}{l}{\textbf{Metrics}}&\multicolumn{1}{l}{\textbf{Test1}}&\multicolumn{1}{l}{\textbf{Test2}}&\multicolumn{1}{l}{\textbf{Test3}}&\multicolumn{1}{l}{\textbf{Avg}.}\\
		\hline 
		\textbf{COVID-19} & Precision& 0.96 &0.76 &0.86  &0.86 \\
		\textbf{Pneumonia}	& Recall&  0.80 & 0.94 & 0.94 &0.89  \\
		& F1-Score& 0.87  &0.84  &0.90  &0.87  \\
		\hline
		\textbf{Bacterial} & Precision& 0.67 & 0.62&  0.53& 0.60\\
		\textbf{Pneumonia}	& Recall& 0.90  &0.54  &0.46  &0.63  \\
		& F1-Score&   0.77&  0.58& 0.49 &0.61  \\
		\hline
		\textbf{Viral}  & Precision&  0.63& 0.71& 0.30 &0.55 \\
		\textbf{Pneumonia} 	& Recall&  0.29 &0.43  &0.30  &0.34  \\
		& F1-Score&   0.39& 0.54 &0.30  &0.41  \\
		\hline
		\textbf{Normal} & Precision&  0.82& 0.93& 0.96 &0.90 \\
		\textbf{Lungs}	& Recall& 0.95  &0.88  &0.93  & 0.92 \\
		& F1-Score&  0.88 &0.90  &0.94  &0.91  \\
		\hline
		\textbf{Overall} & Accuracy&87.99\%&	79.09\%	&79.09\%& 80.31\%\\
		\hline 
	\end{tabular}
	\label{multiClassificationTable}
\end{table}



\subsubsection{Multiclass classification}

\begin{figure}[t]
	\centering
	\includegraphics[width=0.96\linewidth, height= 3.5cm]{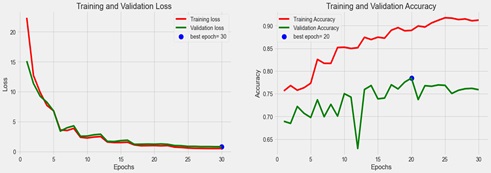}
	\caption{Training and validation loss and accuracy graphs for multi-class pneumonia classification task.}
	\label{trainvsvalidationmulti}
\end{figure}

\begin{figure}[t]
	\centering
	\includegraphics[width=0.98\linewidth, height= 3cm]{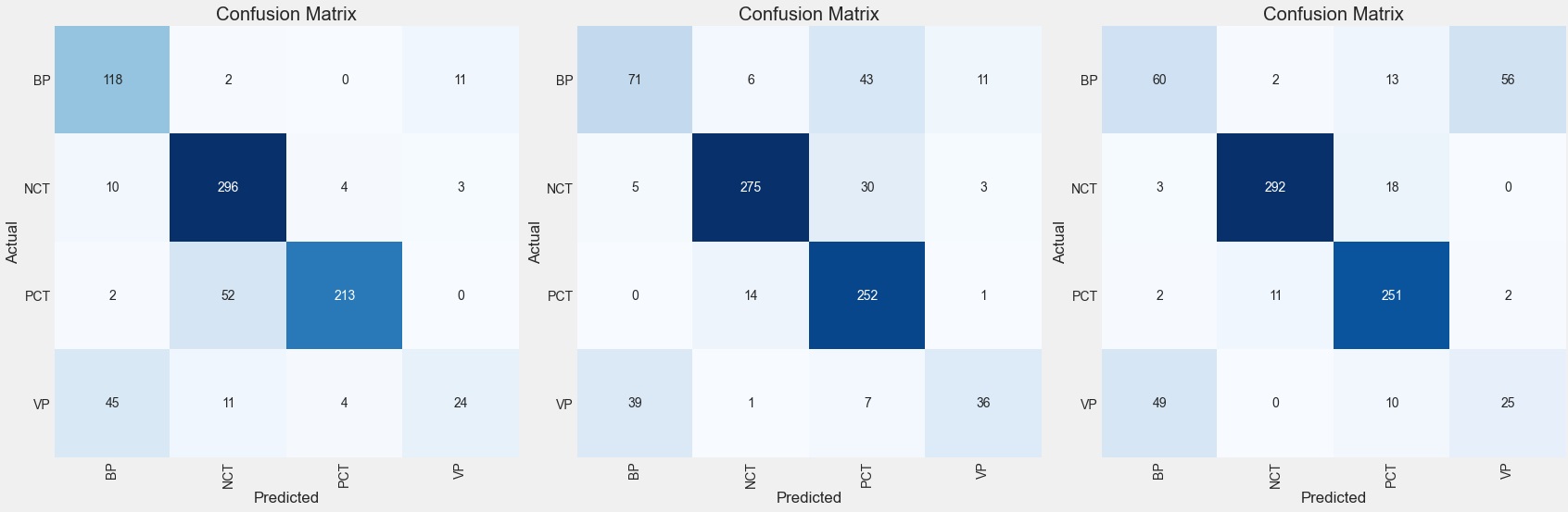}
	\caption{Confusion Matrix of Multiclasses Classification on three test sets [PCT: Positive COVID-19 CT, NCT: Normal Lungs CT, VP: Viral Pneumonia CT, BP: Bacterial Pneumonia].}
	\label{confusionmulti}
\end{figure}

We also performed our experiments on three testing data sets for multiclass classification. The proposed model has detected COVID-19 pneumonia from other pneumonia with an average sensitivity of 0.89, specificity of 0.85, and F1-score of 0.87. The detailed test results for each of the three test sets have been shown in Table~\ref{multiClassificationTable}. 
The confusion matrix for each of these experiments is shown in Fig.~\ref{confusionmulti}. It can be seen from the confusion matrix of test 1 that viral pneumonia is misclassified as bacterial pneumonia. Similarly, COVID-19 has been misclassified a smaller number of times as bacterial pneumonia.
The first test set for multiclass classification shows a sensitivity rate of 0.80 for COVID-19 Pneumonia, 0.90 for bacterial pneumonia, 0.29 for viral pneumonia, and 0.95 for normal lung CT scan. The second test set shows a sensitivity rate of 0.94 for COVID-19 Pneumonia, 0.54 for bacterial pneumonia, 0.43 for viral pneumonia, and 0.88 for normal lung CT scans. The third test set shows a sensitivity rate of 0.94 for COVID-19 Pneumonia, 0.46 for bacterial pneumonia, 0.30 for viral pneumonia, and 0.93 for normal lung CT scan. The training and validation loss graph is shown in Table\ref{trainvsvalidationmulti}. It shows that validation loss is low and slightly higher than the training error, which indicates that the model is a good fit for the data. The early callback function stops the training on epoch 30 when validation loss starts increasing, and it is time to stop.

\subsubsection{Result analysis}

\begin{table}
	\centering
	\footnotesize
		\caption{Comparison between the proposed model and the state-of-the-art models.}
	\begin{tabular}{@{}llll@{}}
		\hline 
		
		\multicolumn{1}{l}{\textbf{Literature}}&\multicolumn{1}{l}{\textbf{Model}}&\multicolumn{1}{l}{\textbf{Class}}&\multicolumn{1}{l}{\textbf{Performance}}\\
		\hline 

		Wang et al.~\cite{wang2020deep}&Modified Inception&	2&	Sensitivity 0.84\\
		& Network&	 &	 \\
		Dastider et al.~\cite{dastider2020rescovnet}&	ResNet152V2&	5&Sensitivity 0.82\\
		Alsharman et al.~\cite{alsharman2020googlenet}&	GoogleNet&	2&	Validation 0.82\\
		Proposed System&InceptionResNet-V2&	4&	Sensitivity \textbf{0.89} \\
		(Multi-class)&	 &	 &	  \\
		Proposed System &	InceptionResNet-V2&	2&	Sensitivity \textbf{0.97}\\
		(binary) &	 &	 &	 \\
		
		\hline 
	\end{tabular}

	\label{compare_models}
\end{table}

As the study on this research is limited, a similar type of work is not available to the best of our knowledge. Most of the research is devoted mostly to binary classification, where CT images are classified into COVID-19 and non-COVID or COVID-19 pneumonia and general pneumonia, not a four-class classification like the one performed in the thesis. Few comparisons are shown in Table~\ref{compare_models}. Their research studies are based on different datasets, mainly using X-ray images. 
Compared with other research studies, this study's classification method has been considered a binary classification of COVID-19 versus other pneumonia. Dastider et al.~\cite{dastider2020rescovnet} classified the combination of X-ray and CT images into COVID-19 and other pneumonia with a sensitivity of 0.82. Wang et al.~\cite{wang2020deep} classified CT images into COVID-19 and viral pneumonia with an accuracy of 79.3\%, and Nesreen et al.~\cite{alsharman2020googlenet} classified CT images with a validation accuracy of 82.14\%. The proposed model surpasses all the compared results in the parameters considered in the classification result.


\section{Limitations and future work}
There are a few limitations of this study that can be overcome in future research, precisely the lack of available datasets and the size of datasets. A larger amount of data is required to overcome the class imbalance challenges. A more interesting approach for future research would be to focus on classifying the degree of severity of the disease to help monitor and treat patients. To make the proposed approach more generalized, we want to utilize it on other larger datasets in different fields like energy, agriculture, and transportation. It is observed that the performance of the model can be improved further by increasing dataset size and using handcrafted features in the future.

\section{Conclusion}
The COVID-19 pandemic affected the entire world's health, economy, and every other life aspect. A sufficient model to identify the differences between similar lung diseases would significantly help diagnose and develop different medications to treat and prevent the development of severe stages of these diseases. This research proposed a novel method to classify different pneumonia types and COVID-19 from CT scan images using deep transfer learning for the CT scan image dataset. The proposed system gives maximum accuracy for binary classification with a sensitivity of 0.97 and multiclass classification with a sensitivity of 0.89. The proposed system can classify COVID-19 from its identical diseases, which can play an essential role in the detection of COVID-19. In the present conditions worldwide, efforts are focused on detecting COVID-19 early to prevent it from spreading. The proposed work can contribute significantly to detecting COVID-19. This finding can be useful to researchers and physicians in making a decision in clinical practice. The proposed system can be employed as a pre-assessment process to reduce the medical staff's workload and to identify and treat the early stages of the disease. The proposed approach can also be applied to process the MRI images as traditional MRI image analysis needs hours of computing time to analyze a vast amount of data of patients. By employing transfer learning, the results are delivered faster. In the medical computer vision field, this approach can speed up the early diagnosis process in diseases such as cancer, tumors, and diabetes. It can be concluded that this approach is helpful for medical practitioners to use as a pre-assessment for the diagnosis of disease. Moreover, it is necessary to develop models competent in distinguishing Covid-19 cases from other similar pneumonia, such as SARS, or even physiological X-rays. Furthermore, it can be simply replicated and implemented in areas with limited equipped medical labs, such as rural areas and developing regions.

\bibliographystyle{ieeetr}
\bibliography{egbib}

\end{document}